\documentclass[aps,prb,twocolumn,groupedaddress,showpacs]{revtex4}
\usepackage{graphicx}

\newcommand{\nn}{\nonumber}
\newcommand{\be}{\begin{equation}}
\newcommand{\ee}{\end{equation}}
\newcommand{\bea}{\begin{eqnarray}}
\newcommand{\eea}{\end{eqnarray}}

\newcommand{\half}{\frac{1}{2}\,}
\newcommand{\la}{\langle}
\renewcommand{\o}{\omega}
\newcommand{\ep}{\epsilon}
\renewcommand{\t}{\theta}
\newcommand{\ra}{\rangle}
\newcommand{\hb}{\hbar}
\renewcommand{\inf}{\infty}

\begin{document}   

\title{Phase phonon spectrum and melting in
a quantum rotor model with diagonal disorder}

\author{W.~A.~Al-Saidi}
\email{Al-Saidi.1@osu.edu}
\author{ D.~ Stroud}
\email{Stroud@mps.ohio-state.edu}
\affiliation{Department of Physics,
The Ohio State University, Columbus, Ohio 43210}
\date{\today}

\begin{abstract}
We study the zero-temperature ($T = 0$) quantum rotor model with
on-site disorder in the charging energy.  Such a model may serve as an
idealized Hamiltonian for an array of Josephson-coupled small
superconducting grains, or superfluid $^4$He in a disordered
environment.  In the approximation of small-amplitude phase
fluctuations, the Hamiltonian maps onto a system of coupled harmonic
oscillators with on-site disorder.  We study the effects of disorder
in this harmonic regime, using the coherent potential approximation
(CPA), obtaining the density of states and the lifetimes of the
spin-wave-like excitations for several choices of the parameters which
characterize the disorder.  Finally, we estimate the parameters
characterizing the $T = 0$ quantum melting of the phase order, using a
suitable Lindemann criterion.
\end{abstract}


\maketitle

\section{Introduction}

The quantum rotor model is a widely studied Hamiltonian which may
serve as an idealized representation of many physical systems.  It
consists of two parts: a potential energy which represents the
coupling between two-component spin variables of fixed length (an
``XY'' Hamiltonian) and the kinetic energy of those spins.  The model
is quantum-mechanical because the potential and kinetic energies do
not commute.

One possible realization of this model is a Josephson junction array,
i.e., a collection of superconducting islands with Josephson or
proximity effect tunneling between the islands.  Such arrays can be
produced experimentally in a wide range of geometries, and with a
great variety of individual junction parameters.  They  may be
described by the quantum rotor model in the limit of small grains.  In
this case the potential energy corresponds to the Josephson coupling,
and the kinetic energy corresponds to the charging energy.  Such
arrays can serve as model systems for studying quantum phase
transitions \cite{sondhi} under conditions such that the experimental
parameters can be readily tuned. \cite{fazio} Recently, it has also
been proposed that Josephson junctions may serve as a quantum bit
(qubit) in quantum information technology, where the quantum logic
operations would be performed by controlling gate voltages or magnetic
fields. \cite{MSS}

The quantum rotor model may also be realized in other physical
systems.  For example, in liquid $^4$He, the potential energy may
represent a discretized form of the gradient energy term in the
Ginzburg-Landau free energy of superfluid $^4$He, and favors a state
in which all parts of the superfluid have the same phase, while the
kinetic energy term is the energy associated with the time-variation
of the order parameter phase.  The model may be particularly useful
for helium films and helium in confined geometries such as porous
glasses.  The spectrum of elementary excitations in such confined
geometries has recently been studied extensively. \cite{sokol} This
spectrum may be treatable by the quantum rotor model.

In this paper, we study a special kind of quantum rotor model, namely,
the two-dimensional model with diagonal disorder in the kinetic
energy.  Our emphasis in this study is to understand what happens to
the spectrum of elementary excitations in the presence of this
disorder.  We also investigate the conditions for the destruction of
phase order in this model arising from such quantum fluctuations in
the phase variables.  This model includes a particularly simple form
of disorder.  In the case of a Josephson junction array (JJA), it
would correspond to fluctuations in the self-capacitance of individual
grains, but no significant randomness in the Josephson coupling
between the superconducting grains.  In the case of superfluid helium,
the diagonal disorder might correspond to randomness in the pore
geometry (for a three-dimensional system) or arising from surface
roughness (for a thin film).

In the present paper, we are primarily interested in the spectrum of
elementary excitations in this model at low temperatures.  We
therefore adopt a simple mean-field approximation which has been
successfully used to treat elementary excitations in other systems.
This is the coherent potential approximation (CPA), as described
originally by Soven, \cite{soven} Taylor {\it et
al},\cite{taylor2,taylor1} and Velick\'{y} {\it et al.}\cite{velicky}
For the present problem, the CPA may be adequate at very low
temperatures, under conditions such that the original Hamiltonian can
be approximated as harmonic.  At higher temperatures, or under
conditions such that the harmonic approximation is inadequate, it will
be necessary to consider a more sophisticated approximation, such as
the self-consistent harmonic approximation
(SCHA), \cite{simanek,wood,chakra} or a field-theoretic treatment in
which the charging energy is mapped onto a coupling in the imaginary
time direction. \cite{fisher}

The remainder of this paper is organized as follows.  In the next
section, we present the model Hamiltonian, and its relation to a set
of coupled harmonic oscillators.  In Sec. III, we describe how the CPA
may be used to study the disorder-averaged properties of the model
Hamiltonian in the harmonic approximation.  Our numerical results based
on the CPA are presented in Sec. IV, followed by a concluding
discussion in Sec. V.

\section{Model Hamiltonian}

We will consider the following model Hamiltonian:
\be 
H=\half \sum_{j} U_j \,{n_j}^2 + \sum_{j k }J_{jk}[1- \cos(\theta_j-
\theta_k )].  \label{eq:oham}
\ee 
Here $\theta_j$ is the phase of the quantum rotor, and $n_j$
is a variable which is quantum-mechanically conjugate to the
angle $\theta_j$. $n_j$ and $\theta_k$ are assumed to
have commutation relations
\begin{equation}
[n_j, \theta_k] = -i\delta_{jk},
\label{eq:comm}
\end{equation}
where $\delta_{jk}$ is a Kronecker delta function and $[\ldots]$
represents a commutator.  In Eq.\ (\ref{eq:oham}), the first
sum runs over sites $j$, and 
the second sum runs over all distinct pairs of sites $j$ and $k$. 
If this model is assumed to describe a Josephson junction
array, then the first term represents the capacitive energy of the 
array in the diagonal approximation.  The most general form of this
capacitive energy would be $\frac{1}{2}\sum_{jk} U_{jk}n_jn_k$;
in Eq.\ (\ref{eq:oham}), we make the approximation 
$U_{jk} = U_j\delta_{jk}$, where $U_j$ is related to the
diagonal capacitance $C_j$, or alternatively to 
an effective ``mass'' $M_j$, 
\begin{equation}
U_j = \frac{4e^2}{C_j} \equiv \frac{\hb^2}{M_j}.  
\end{equation}
The second term represents the sum of the Josephson
coupling energies between grains $j$ and $k$.  In Eq.\ 
(\ref{eq:oham}),
it is implicitly assumed that there is no contribution to $H$
arising from dissipation.  In this paper, we shall assume
that $J_{ij} \neq 0$ only if $i$ and $j$ are nearest-neighbor
grains, and that, furthermore, there is no disorder in this
nearest-neighbor coupling, i.e., that $J_{ij} = J$ for all
nearest neighbor pairs $i$ and $j$.  Hereafter, we shall refer
to this model Hamiltonian as describing a Josephson junction
array, although, as noted above, the model is also applicable,
in principle, to other physical systems.

For sufficiently low temperatures, we assume that
the above Hamiltonian
can be approximated by its harmonic form:
\be
H_{\mathrm{harm}} = \half \sum_j  U_j\, {n_{j}}^{2} +\half J\,\sum_{\la  j k \ra}(\theta_j
-\theta_k)^2  \equiv K + V,
\label{eq:ham}
\ee
where the second sum runs over distinct nearest neighbor pairs.
While one can go beyond the harmonic approximation by using
a self-consistent phase phonon approach, \cite{simanek,wood} we
shall here consider only the harmonic approximation, which is
more easily combined with a treatment of disorder.
The Heisenberg equation of motion of the phase variable is 
$i \hb \,\dot{\t_j}= [\t_j,H]=i U_j n_j$, where the last equality
follows from the commutation relation (\ref{eq:comm}).
This equation makes clear that $H_{\mathrm{harm}}$ is formally analogous
to an array of ``masses'' $M_j$ harmonically coupled by 
springs of ``spring constant'' $J$.  However, in contrast to 
a real lattice of masses and springs, there is no polarization
degree of freedom.  $K$ and $V$ represent
the kinetic and potential energy terms in this Hamiltonian.
A disordered array of such oscillators can thus be treated
by the same methods used to treat disordered arrays of coupled
oscillators.

In this paper, we shall consider a bimodal distribution of the
$U_j$'s.  Thus we take $U_j = U_1$ or $U_2$ with probabilities $p$ and
$1 - p$ respectively.  We also assume that the type $1$ and type $2$
grains are randomly distributed on an ideal lattice, with no
correlation between the occupancies of neighboring sites.  With this
choice, the Hamiltonian (\ref{eq:ham}) is analogous to the classic
problem of substitutional mass disorder in phonon spectra, except that
the ``phase phonon'' excitations have only one possible polarization.
This problem has previously been studied by Soven, \cite{soven}
Taylor\cite{taylor2}, and Elliott and Taylor, \cite{taylor1} using the
coherent potential approximation. \cite{soven,taylor2,taylor1,velicky}
In the present paper, we shall use the same approach to study the
effects of disorder in the ``phase phonon'' spectrum.  Note that
because the entire Hamiltonian is invariant under a global rotation of
phases, the long-wavelength phase phonons are Goldstone bosons with
frequencies that go to zero linearly in the wave vector ${\bf k}$, even in the
presence of disorder, and even in two dimensions.

\section{Coherent potential approximation}

\subsection{General equations}

The CPA provides an approximate way of calculating the 
disorder-averaged Green's function
\begin{equation}
\bar{G}(z) \equiv \left\langle \frac{1}{z - H_{\mathrm{harm}}} \right\rangle_{\mathrm{dis}},
\end{equation}
where $z$ is a complex variable, $H_{{\mathrm{harm}}}$ is the Hamiltonian
operator, and $\langle...\rangle_{\mathrm{dis}}$ represent an average over
disorder realizations.  Of particular interest is the diagonal
matrix element of this operator, which we write
\begin{equation}
F(z) = \langle j |\bar{G}(z)| j \rangle \equiv \bar{G}(j, j; z).
\label{eq:fdef}
\end{equation}
Because the operator $\bar{G}$ is disorder-averaged, 
$F(z)$ is independent of $j$.  

Also, since $\bar{G}(z)$ is disorder-averaged, it can be expressed
in terms of a periodic operator $H_{\mathrm{eff}}$ by the relation
\begin{equation}
\bar{G}(z) = \frac{1}{z - H_{\mathrm{eff}}}.
\end{equation}
For the present case of ``site-diagonal'' disorder, $H_{\mathrm{eff}}$
can be written
\begin{equation}
H_{\mathrm{eff}} = K_{\mathrm{eff}} + V,
\end{equation}
where $K_{\mathrm{eff}}$ can be expressed as
\begin{equation}
K_{\mathrm{eff}} =\frac{\hbar^2}
{2M_{\mathrm{eff}}}\, \sum_j n_j^{2}.
\end{equation}
Here $M_{\mathrm{eff}}(z)$ is a complex, frequency-dependent ``effective
mass'' which is, however, the same for each site $j$.
It is convenient to express $M_{\mathrm{eff}}$ in terms of two other
quantities - the ``virtual-crystal'' mass
\begin{equation}
M_{\mathrm{VCA}} = pM_1 + (1-p)M_2,
\end{equation}
and a dimensionless, complex self-energy  function
$\Sigma(z)$ defined by
\begin{equation}
M_{\mathrm{eff}}(z) = M_{\mathrm{VCA}}\left[1 - \Sigma(z)\right].
\end{equation}

In the CPA, it is found that $F(z^2)$ and $\Sigma(z)$ are 
connected by the complex scalar equation \cite{diamond}
\begin{equation}
M_{\mathrm{VCA}} \Sigma(z)= \frac{ p(1-p) (\delta M)^2 z^2 F(z^2)}
	{ 1+z^2\left[(1-2p)\delta M + M_{\mathrm{VCA}} 
\Sigma(z)\right]F(z^2)}
\label{eq:selfenergy}.
\end{equation}
Here $\delta M \equiv M_1 -M_2 \equiv -\epsilon\, M_2 $, where
$\epsilon \equiv 1 - M_1/M_2 = 1 - C_1/C_2$, and we have
adopted the convention that $C_1 < C_2$. 

In order to complete the approximation, we need to express $F(z^2)$ in
terms of $\Sigma(z)$.  This is readily accomplished as follows.
First, we consider the analogous virtual-crystal Green's function
$G_{\mathrm{VCA}}(z) \equiv 1/(z - K_{\mathrm{VCA}} - V)$, where
$K_{\mathrm{VCA}} = \sum_j n_j^2/[2M_{\mathrm{VCA}}]$.
Since $K_{\mathrm{VCA}} + V$ is periodic, $G_{\mathrm{VCA}}(z^2)$ is
diagonal in momentum space. In particular, the diagonal matrix element
is given by
\begin{equation}
\langle {\bf q}|G_{\mathrm{VCA}}(z^2)|{\bf q}\rangle
= \frac{1}{z^2 - \omega_{\bf q}^2} \equiv G_{\mathrm{VCA}}({\bf q}; z^2),
\end{equation}
where $\omega_{\bf q}$ is the frequency of a ``phase phonon''
of wave vector ${\bf q}$ in the virtual crystal, and 
$|{\bf q}\rangle$ and $|i\rangle$ are related by
$|{\bf q}\rangle = N^{-1/2}\sum_ie^{i{\bf q}\cdot {\bf R}_i}|i\rangle$,
${\bf R}_i$ being the position of the $i$th grain and
$N$ being the number of grains in the lattice (which may be
assumed to have periodic boundary conditions).  The corresponding
diagonal matrix element $F_{\mathrm{VCA}}(z^2)$ is given by
\begin{equation}
F_{\mathrm{VCA}}(z^2) = \frac{1}{2 N M_{\mathrm{VCA}}}\sum_{\bf q}G_{\mathrm{VCA}}({\bf q}, z^2),
\end{equation}
where the sum runs over ${\bf q}$'s in the first Brillouin zone
of the grain lattice.  Next, we introduce the virtual-crystal
density of states $g_{\mathrm{VCA}}(\eta)$ by
\begin{equation}
g_{\mathrm{VCA}}(\eta) = \frac{1}{2N}\sum_{\bf q}\delta(\eta - \omega_{\bf q}),
\end{equation}
in terms of which $F_{\mathrm{VCA}}(z^2)$ takes the form
\begin{equation}
F_{\mathrm{VCA}}(z^2)
= \frac{1}{M_{\mathrm{VCA}}} \int_0^\infty \frac{g_{\mathrm{VCA}}(\eta)}
{z^2 - \eta^2}d\eta.
\label{eq:fvca}
\end{equation}

Finally, we can obtain $F(z^2)$ for the actual disordered
crystal in terms of $F_{\mathrm{VCA}}(z^2)$ simply by making the
replacement $z^2 \rightarrow z^2\left[1 - \Sigma(z)\right]$,
so that
\bea
F(z^2) &=& F_{\mathrm{VCA}}\left(z^2\left[1 -
\Sigma(z)\right]\right)\nn \\
&=&\frac{1}{M_{\mathrm{VCA}}} \int_0^\infty
\frac{g_{\mathrm{VCA}}(\eta)}{z^2\left[1 - 
\Sigma(z)\right] - \eta^2}d\eta.
\label{eq:fz}
\eea
Equations.\ (\ref{eq:selfenergy}) and (\ref{eq:fz}) represent two
coupled equations for $F(z^2)$ and $\Sigma(z)$ which can be
solved simultaneously, given a model for $g_{\mathrm{VCA}}(\eta)$.

In the present paper, we will consider a two-dimensional (2D)
lattice with a simple Debye-like density of states
defined by
\be
g_{\mathrm{VCA}}(\eta)= 
\frac{\eta}{ 2 \pi v^2 }  \,\Theta(\omega_D-\eta)
\label{eq:densityVCA}.
\ee
Here $v=\sqrt{J/ M_{\mathrm{VCA}}}$ is the speed of
the ``phase phonon'' excitations in the virtual crystal,
and $\Theta(\omega_D-\eta)$ is a unit step function.  
The Debye cutoff frequency
\be
\omega_D=(4 \pi v^2 )^{1/2}= \left( 4 \pi J/M_{\mathrm{VCA}}
\right)^{1/2} \label{eq:omegaD}, 
\ee
follows from requiring
that the 
number of phase phonon states be normalized to one per grain.
The quantity $F_{\mathrm{VCA}}(z^2)$ follows readily from Eq.\ (\ref{eq:fvca}):
\be
F_{\mathrm{VCA}}(z^2) = \frac{1}{M_{\mathrm{VCA}}\omega_D^2}
\ln\left[\frac{z^2}{z^2 - \omega_D^2}\right].
\label{eq:fdebye}
\ee
Equations (\ref{eq:selfenergy}), 
(\ref{eq:fz}), and (\ref{eq:fdebye}) 
constitute a self-consistent set of equations
which can now be explicitly solved to give the properties
of the phase excitations in the disordered system,
within the CPA.

\subsection{Lattice properties}

\subsubsection{Density of states}

The density of phase phonon excitations within the CPA is
given by \cite{diamond}
\be
g_{\mathrm{CPA}}(\omega)=-\frac{2}{\pi \omega}{\mathrm{Im}} \int_0^\infty 
 \frac{\eta^2\, g_{\mathrm{VCA}}(\eta)}{\omega^2[1-\Sigma(\omega)]-\eta^2}
d\eta.
\label{eq:densityCPA}
\ee 
Using Eq.\ (\ref{eq:densityVCA}) for $g_{\mathrm{VCA}}(\eta)$, this
expression can be further simplified to 
\be
g_{\mathrm{CPA}}(\omega)= -\frac{2}{\pi \omega} {\mathrm{Im}}[ h(x^2) ],
\label{eq:density} 
\ee
where
$x^2 =(\omega^2/\omega_D^2)\left[1-\Sigma(\omega)\right]$
and $h(x^2)= x^2\ln\left[x^2/(x^2-1)\right] - 1$.
Thus, once we have obtained the self-energy from
Eq.\ (\ref{eq:selfenergy}), we can
find the density of vibrational states, $g_{\mathrm{CPA}}(\omega)$
from Eq. (\ref{eq:density}). \cite{note}

The CPA can also be used to give the component densities
of states $g^{(\alpha)}_{\mathrm{CPA}}(\omega)$ ($\alpha$ = 1, 2), defined to be
the average density of phase phonon density of states on grains
of type $\alpha$.  This is obtained in terms of the average 
diagonal matrix element $\bar{G}^{(\alpha)}(j, j; z^2)$ for a grain of 
type $\alpha$.  In the CPA, this matrix element takes
the form: \cite{velicky}
\be
\bar{G}^\alpha (j,j; z^2)= \frac{F(z^2)}
{1- (M_{\mathrm{VCA}} - M_\alpha  - M_{\mathrm{VCA}}\Sigma)\,z^2\,F(z^2)},
\label{eq:gd}
\ee
where $F(z^2)$ is defined in Eq.\ (\ref{eq:fdef}). 
The average density of states at a grain of type $\alpha$ 
(normalized so as to integrate to unity over positive frequencies)
is given by
\begin{widetext}
\be
g^{(\alpha)}_{\mathrm{CPA}}(\omega)=-\frac{2}{\pi \omega}{\mathrm Im}
\left[\frac{ h ( x^2) }{ 1- (M_{\mathrm{VCA}} - M_\alpha  -
M_{\mathrm{VCA}}\Sigma)  
\omega^2\,F(\omega^2) } \right]\label{eq:densityD},
\ee
where $x^2$ and $h(x^2)$ were defined below Eq. (\ref{eq:density}).

Within the CPA, the component densities of states have the
pleasing property
\begin{equation}
p g_{\mathrm{CPA}}^{(1)}(\omega) + (1-p)g_{\mathrm{CPA}}^{(2)}(\omega)
= g_{\mathrm{CPA}}(\omega).
\label{eq:densityA}
\end{equation}
Thus, the total phase phonon density of states equals the
sum of the properly weighted partial densities of states
on the two species of grains.
In our calculations, we have confirmed that this property is,
indeed, satisfied for our particular choice of the  Debye model.

\subsubsection{Spectral Function}

Our self-consistent set of equations permits calculation of another useful
quantity: the spectral function, defined by
\be
{\mathrm Im} \bar{G}({\bf q},\o^2)= 
\frac{\omega^2 {\mathrm Im} \left[ \Sigma(\omega^2)\right]}
     {\left\{\omega^2 (1- \mathrm{Re}\left[\Sigma(\omega^2)\right]) 
				- {\omega_{{\bf q}}}^{2}\right\}^2 
	+ 
	\left\{\omega^2\mathrm{Im}
				\left[\Sigma(\omega^2)\right]\right\}^2}.
	\label{eq:SF}
\ee
\end{widetext}
The spectral function gives the frequency distribution of
excitations of wave number ${\bf q}$ in the array, and, for
our Debye model, depends only on the magnitude $q$.  
The full width at half maximum (FWHM) of the spectral function is 
inversely proportional to the decay time of the spin-wave-like 
excitations at wave vector ${\bf q}$.

\subsubsection{Mean-Square Phase Fluctuations}

Finally, we can use our CPA calculations to infer the mean-square
phase fluctuations at the $j${th} grain.  We denote this quantity
$\langle |\theta_j(t)|^2\rangle$, where $\langle \ldots \rangle$
denotes a quantum-mechanical average, and $\theta_j(t)$ denotes the
phase of the $j${th} grain at time $t$.  A quantum-mechanical average
is appropriate at temperature $T = 0$.  At finite temperatures, one
should carry out both a quantum and a thermal average.  However, such
an average diverges at $T \neq 0$ in two dimensions ($d = 2$), because
of contributions from the long-wavelength phase phonons.  In the
present discussion, therefore, we discuss only the $T = 0$ limit.

We will consider the quantity $\langle |\theta_j(t)|^2 \rangle$ for a
lattice with Josephson coupling $J$ and self-capacitance
$C_j$, within the harmonic approximation.  In this case, one can
simply adapt the discussion of Ref.\ \onlinecite{rmp} for lattice
vibrations in mass-disordered systems.  In fact, the present
problem is slightly easier since there are no polarization degrees
of freedom.  To obtain this quantity, we first use
a result from Ref.\ \onlinecite{rmp} that 
\be
\la \theta_j(t) \theta_k(0) \ra = - \frac{\hb}{\pi}{\mathrm{Im}}
				\int_{-\inf}^{\inf}  e^{i \omega t}
				G(j,k;\o^2) d\o\label{eq:thij},
\ee
where, at $T = 0$, 
$G(j, k; \o^2)= \langle j|\frac{1}{\omega^2 - H}|k\rangle$.
The desired quantity is actually 
$\lim_{t \rightarrow 0}\la \theta_j(t)\theta_j(0)\ra$
averaged over disorder realizations, which we write
\be
\la |\t_j|^2 \ra_{\mathrm{dis}}= -\frac{\hb}{\pi}{\mathrm Im}\int_{0}^{\inf}
F(\o^2) \, d\o \label{eq:thj2},
\ee
where we have used the definition of
$F(z^2) $ in Eq. (\ref{eq:fdef}).   The
operators $\t_j$ are computed at times $t = 0$ (though this
average is time-independent).

We first consider the virtual crystal lattice, in which 
$C_j = C_{\mathrm{VCA}}$ for all $j$.  In this case,
$ \la |\t_j|^2 \ra$ is readily calculated from Eq.\
(\ref{eq:thj2}), using
$F_{\mathrm{VCA}}(z)$ from Eq.\  (\ref{eq:fdebye}), with the result
\be
\la |\t_j|^2 \ra_{\mathrm{VCA}}= \frac{\hbar}{ \omega_{D} M_{\mathrm{VCA}}}\label{eq:dw}
\ee
where  $\o_{D}= \sqrt{4 \pi J/M_{\mathrm{VCA}}} 
= \sqrt{16 \pi e^2 J/(\hb^2\,C_{\mathrm{VCA}})}$.
 
For the actual disordered lattice, we can calculate within
the CPA not
only the full disorder average of $\la |\theta_j|^2\ra_{\mathrm{dis}}$, 
but also the disorder average of
$\la |\theta_j|^2\ra$ over sites of type $\alpha$,
which we denote $\la |\t_j|^2\ra^{(\alpha)}$.
The result is
\begin{widetext}
\bea
 \la |\t_j|^2 \ra^{(\alpha)} &=&
	\frac{-\hb}{\pi} {\mathrm Im}\int_{0}^{\inf}
			\bar{G}^\alpha(j,j;\o^2) \, d\o\\
		             &=&
        \frac{-\hb}{\pi}{\mathrm Im} \int_{0}^{\inf}
	\frac{F(x^2)}{1-(M_{\mathrm{VCA}}-M_\alpha-M_{\mathrm{VCA}}\Sigma)
						\o^2 F(x^2)}d\o,
\eea
\end{widetext}
where $x^2=(\omega^2/\omega_D^2)\left[1-\Sigma(\omega)\right]$.
The full $\langle |\theta_j|^2\rangle_{\mathrm{dis}}(p)$  in the CPA 
satisfies
\begin{equation}
\langle|\theta_j|^2\rangle_{\mathrm{dis}}(p) = \sum_\alpha p_\alpha\langle|\theta_j|^2
\rangle^{(\alpha)},
\end{equation}
where the sum runs over the two species $\alpha$, and in
our notation
$p_1 = p$, and $p_2 = 1-p$.

We can use these equations to compute various ratios,
such as $\langle|\t_j|^2\rangle_{\mathrm{dis}}(p)/\langle|\t_j|^2\rangle_{\mathrm{VCA}}$,
and hence to see how the mean-square phase fluctuations increase
with increasing disorder.

\begin{figure}
\vspace{0.5in}
\includegraphics[width=9cm]{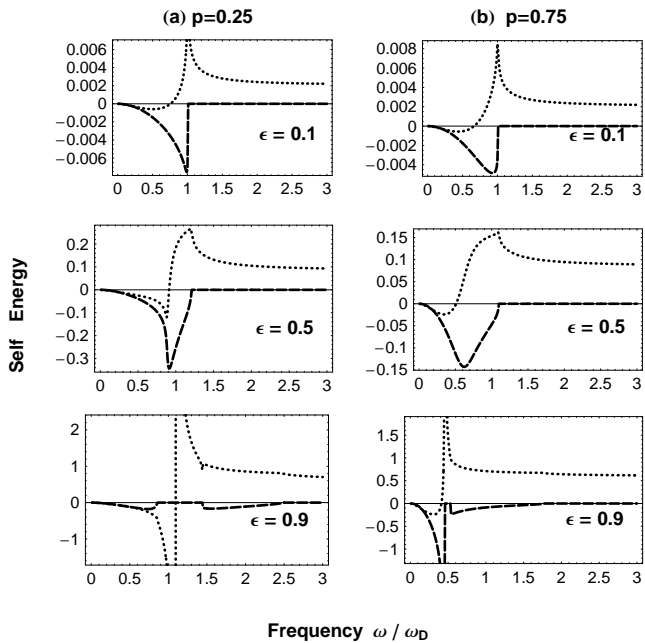}
\caption{
\label{fig1}
 (a) The real (dotted line) and imaginary (dashed line) parts of
the self-energy, $\Sigma(\o)$, plotted as a function of the scaled
frequency $\o/\o_D$ for three different cases: $\ep= 0.1, 0.5,$ and $
0.9 $, at a concentration $p=0.25$ of light defects.  Note that
$\omega_D$ is a function of $p$ and $\epsilon$
[cf. Eq. (\ref{eq:omegaD})]. (b) Same as (a)
except that $p=0.75$.  Note the different vertical scales on different portions
of the Figure.}
\end{figure}

\begin{figure}[t]
\vspace{0.5in}
\includegraphics[width=9cm]{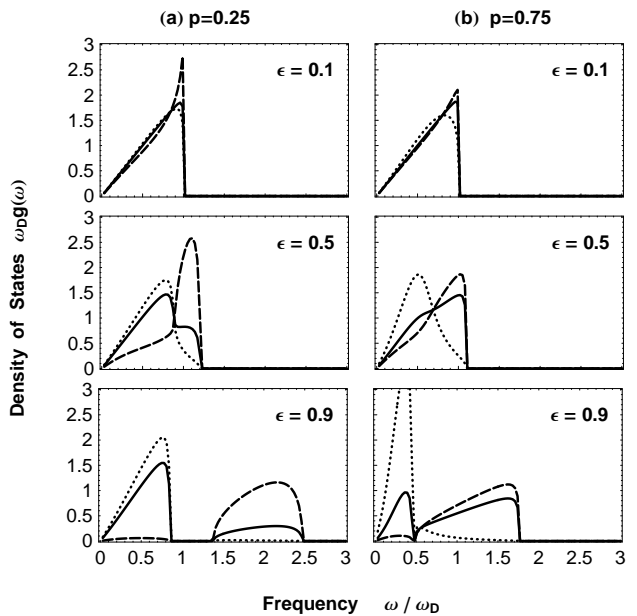}
\caption{
\label{fig2}
(a) Normalized density of states of the phase phonon
excitations, plotted as a function of the scaled frequency $\o/\o_D$
calculated for a concentration $p=0.25$ of low-capacitance defects and
three different values of the disorder parameter $\ep=1-C_1/C_2$, as
indicated in the Figure.  The total density of states for phase phonon
excitations, $g_{\mathrm{CPA}}(\o)$, is the solid line, while the
component density of states $g^{(1)}_{\mathrm{CPA}}(\o)$ and
$g^{(2)}_{\mathrm{CPA}}(\o)$ associated with low-capacitance and
high-capacitance grains are the dashed and dotted lines, respectively. The
concentration of light defects (type $1$) is $p=0.25$. (b) same as (a) except
that $p=0.75$. As in Fig.1, $\omega_D$ is a function of both $p$ and
$\epsilon$ [Eq. (\ref{eq:omegaD})].}
\end{figure}

\section{Numerical Results}

We have solved the model described in the preceding section to obtain
the self-energy, density of states, spectral function, and mean-square
phase fluctuations of a two-dimensional lattice, using a Debye phase
phonon spectrum.  The numerical solution are obtained from
straightforward iterative solution of the three self-consistent
equations (\ref{eq:selfenergy}), (\ref{eq:fz}), and (\ref{eq:fdebye}).
It is important to take care that the resulting solutions are the
proper physical ones, with $\mathrm{Im}\,\Sigma(\omega) \leq 0$ and with
both $F(z^2)$ and $\Sigma(z)$ varying smoothly with $z$.

In Fig.\ 1 we show the real and imaginary parts of the self-energy,
$\Sigma(\o)$ (dotted and dashed curves respectively), plotted against
the scaled frequency $\o/\o_D$, where $\o_D= (4 \pi J/
M_{\mathrm{VCA}})^{1/2}$ is the virtual crystal Debye frequency.  We
have carried out our calculations for two concentrations of the
low-mass defects (type $1$), $p=0.25$ and $p=0.75$; the results are shown in
Figs.\ 1(a) and 1(b). The three different panels correspond to
different values of the ratio $\ep \equiv 1 - C_1/C_2 = 0.1, 0.5,$ and
$ 0.9 $.  These ratios range from small deviations in capacitance between
type $1$ and type $2$ ($\ep = 0.1$) to large deviations ($\ep = 0.9$).
Note the differences in the scales of the plots.

Given the self-energy, several other physical quantities can be
calculated, as explained in Sec. III.  For example, the averaged
density of states of phase-phonon excitations can be obtained within
the CPA from Eq.\ (\ref{eq:density}), while the averaged component
densities of states follow from Eq.\ (\ref{eq:densityD}).  In Fig.\ 2
we show the average density of states $g_{\mathrm{CPA}}(\omega)$
obtained within the CPA (solid line), as well as the component
densities of states $g^{(1)}_{\mathrm{CPA}}(\o)$ (dashed line) and
$g^{(2)}_{\mathrm{CPA}}(\o)$ (dotted line).  The sum rule given in
Eq.\ (\ref{eq:densityA}) is evidently satisfied by these three
densities of states.

\begin{figure}[t]
\vspace{0.5in}
\includegraphics[width=9cm]{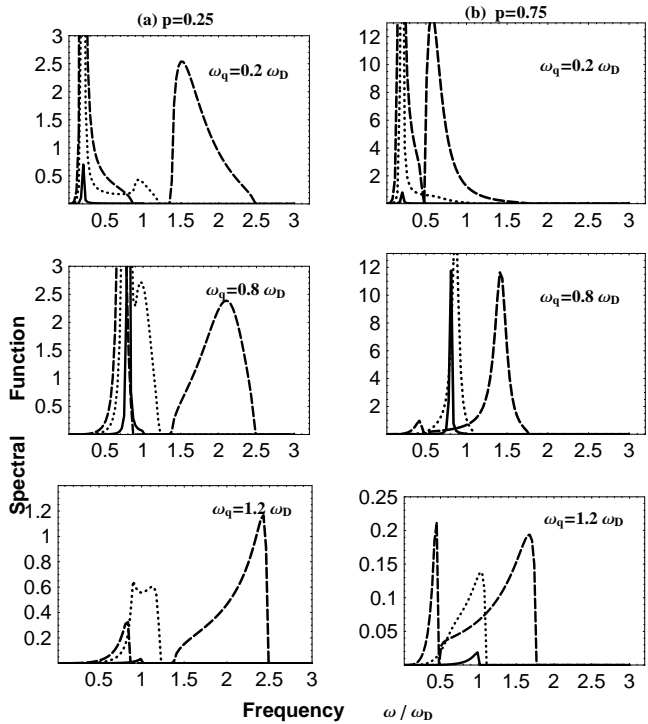}
\caption{
\label{fig3}
(a) Spectral function ${\mathrm Im}\bar{G}({\bf q},\o^2)$, plotted
against the scaled frequency $\o/\o_D$ for three different mode
frequencies $\omega_{\bf q}$ as indicated in each panel. The
concentration of light defects is $p=0.25$.  For each $\omega_{\bf q}$
we have considered $\ep \equiv 1 - C_1/C_2= 0.1$ (solid line), $0.5$
(dotted line), and $0.9$ (dashed line), corresponding to increasing amounts
of disorder.  For $\o_{\bf q}/\o_D < 1 $, the spectral function always
exhibits a single sharp peaked centered at $\o=\o_{\bf q}$. (b) same
as (a) except that $p=0.75$.}
\end{figure}

These results show characteristic features expected from CPA
calculations.  For the case where the capacitance of type $2$ is close
to that of type $1$ ($\ep=0.1$), the full and the two component
densities of states are all very similar, and all resemble the
virtual-crystal result.  As $\ep$ increases, there start to be more
conspicuous differences between the partial densities of states on the
light-mass (type $1$) and heavy-mass (type $2$) grains.  
In particular, there is a clearly-developed band gap between the two
classes of states for $\ep=0.9$. This behavior is well known as the
``split-band'' regime in the phonon problem.  The light-mass phase
phonons presumably correspond to localized modes, though this
localization is not probed in the CPA.

In Fig.\ 3, we plot the spectral function defined in Eq.\
(\ref{eq:SF}) as a function of the scaled frequency $\o/\o_D$ for
several choices of parameters.  Our calculations are carried out at
three different mode frequencies, $\o_{\bf q}/\o_D= 0.2$, $0.8$, and
$1.2$, which are shown in the three different panels, and for two
different defect concentrations, $p = 0.25$ and $p = 0.75$, which are
shown in Figs.\ 3(a) and 3(b).  In each of these cases, we have also
considered three values of the mass ratio parameter: $\ep= 0.1, 0.5,$
and $0.9$.  (Note in particular the vertical scale changes in the Figures.)
For $\o_{\bf q}/\o_D < 1$, the spectral functions in all cases exhibit
single sharp delta-function-like peaks centered at $\o=\o_{\bf q}$.
However, for $\o_{\bf q}/\o_D > 1 $, there is some spectral strength
which shows up as weak peaks at higher frequencies; these are
presumably due to the localized modes in the light-mass bands. Note
also that for $\ep=0.1$ there is almost no spectral weight for
$\o_{\bf q}/\o_D > 1 $.

\begin{figure}[t]
\vspace{0.5in}
\includegraphics[width=8.3cm]{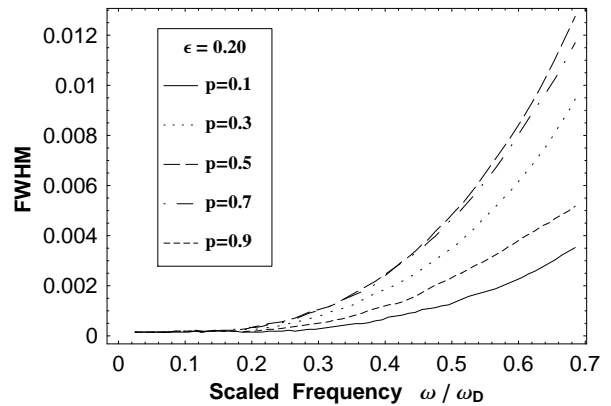}
\caption{
\label{fig4}
The full width at half maximum (FWHM) of the spectral function,
plotted versus scaled frequency $\o/\o_D$ for $\ep = 0.2 $, and for
several concentrations $p$ of the small-capacitance grains, as
indicated in the legend.}
\end{figure}

\begin{figure*}
\vspace{0.5in}
\includegraphics[width=16.3cm]{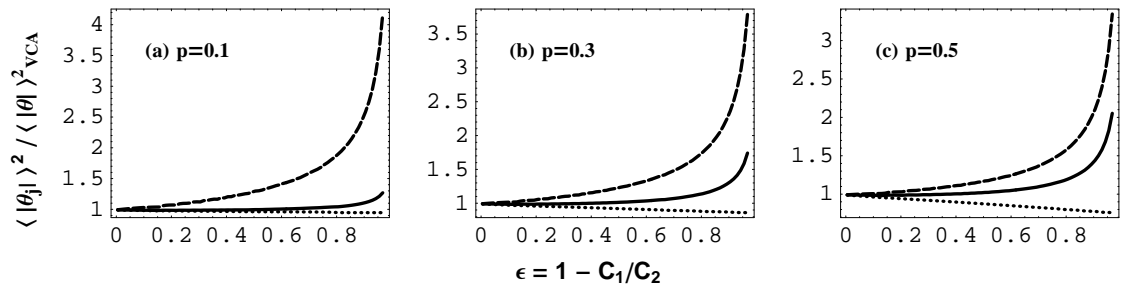}
\caption{
\label{fig5}
 (a). Mean-square phase fluctuation $\langle
|\theta_j|^2\ra_{\mathrm{dis}}/ \langle
|\theta_j|^2\ra_{\mathrm{VCA}}$, as evaluated in the CPA, plotted as a
function of $\ep= 1-C_1/C_2 $ for $p=0.1$. The solid line shows the
CPA average value, the dashed line shows the mean-square phase
fluctuations on a type $1$ grain, and the dotted line on a type $2$
grain.  (b). same as (a) except that $p=0.3$. (c). same as (a) except
that $p=0.5$.}
\end{figure*}

In each of the cases of Fig.\ 3, we have also calculated the full
width of the spectral function peak at half-maximum (FWHM).  This
width is inversely proportional to the lifetime of the phase phonon
like excitations. In Fig.\ 4, we plot this FWHM as a function of
$\o_{\bf q}/\o_D$ for $\ep = 0.2 $, and for several concentrations of
the light defects as indicated in the legend.  Evidently, for a given
$\omega_{\bf q}$, the excitation lifetimes decrease, as expected, as
$p$ increases from $0$ to $0.5$, then increases again as $p$ increases
from $0.5$ to $1$.  This behavior is consistent with the well-known
result of second-order perturbation theory that this lifetime should
vary approximately as $1/[p(1-p)]$ in the weak-disorder
regime. \cite{velicky}

Finally, we turn to the transition from phase coherence to phase
incoherence in this system.  If the model Hamiltonian is assumed to
represent an array of small Josephson junctions, this corresponds to a
superconductor-insulator (S/I) transition.  It is believed that
superconductivity in Josephson junction arrays is destroyed when the
ratio of the charging energy to the Josephson coupling energy exceeds
some characteristic limiting value. \cite{fazio} In fact, the S/I
transition is a standard example of a $T = 0$ quantum phase
transition. \cite{sondhi}  In two-dimensional square arrays with
nearest-neighbor coupling, it has been found experimentally that the
S/I transition occurs near $U/J = 1.7 \equiv
(U/J)_{\mathrm{cr}}$. \cite{fazio}

In our discussion, we will assume a simple Lindemann melting
criterion, namely, that superconductivity is destroyed when the mean
fluctuations in the phase, given in Eq. (\ref{eq:thj2}), exceeds some
limit to be determined below. (In $d = 2$, such a criterion can be
plausible only at $T = 0$, 
since the mean-square phase fluctuation
diverges at any finite $T$.)  If all the grains have the same charging
energy, then the mean-square fluctuation is given by Eq.\
(\ref{eq:dw}), which implies a critical value of
\be
\la |\theta_j|^2 \ra_{\mathrm{cr}} 
= \left({U}/{J}\right)^{1/2}_{\mathrm{cr}} \approx 1.3 \label{eq:crit}.
\ee 
Here $j$ can be any lattice site, and we have used
$\left({U}/{J}\right)_{\mathrm{cr}}=1.7$.

How can this Lindemann criterion be extended to a lattice with
diagonal disorder in the charging energy?  We speculate that a
modified Lindemann criterion may still be usable, in the following
way.  Let us consider the model studied in this paper, in which there
are two types of grains: ``light'' and ``heavy'' (or small and large
capacitance).  We suggest that a possible Lindemann criterion for
destruction of superconductivity in this case is that the mean-square
phase fluctuations, averaged over the two types of grains, 
should exceed the same critical value as in the ordered case.

In Fig.\ 5 we plot the calculated ratio $\la
|\theta_i|^2\ra^{({\alpha})}/\la|\theta_i|^2\ra_{\mathrm{VCA}}$, which
gives the mean-square phase fluctuations on grains of type $\alpha =
1, 2$ with respect to those of the virtual crystal (dashed and dotted
lines).  We also show
$\la|\theta_i|^2\ra_{\mathrm{dis}}/\la|\theta_i|^2\ra_{\mathrm{VCA}}$
which represents the average for the entire lattice (solid line).  All
quantities are calculated in the CPA as a function of $\ep= 1-C_1/C_2
$ for several different values of $p$, as indicated in the Figure.  As
expected, and as is clear from the Figure, for all values of $p$, $\la
|\theta_i|^2\ra^{(1)} > \la |\theta_i|^2\ra^{(2)}$ because component
$1$ represents the light defects.

In principle, one might make an estimate of the critical value of
$(U/J)^{1/2}$ at which melting occurs, as a function of $p$ and
$\epsilon$, by assuming that melting occurs when $\la|\theta_i|^2\ra
\ge 1.3$, as in the ordered lattice.  Thus, given the properties of
the lattice, one can easily extract an estimate of the melting
parameters from this Figure.

\section{Summary}

In this paper we have considered the behavior of a 2D quantum rotor
model with diagonal disorder.  In the case of Josephson junction
arrays, this model corresponds to an array of small Josephson
junctions with diagonal capacitive disorder.  We have used the CPA to
estimate the effects of disorder on the phase phonon density of
states, the spectral functions, and the self-energy, all within the
harmonic approximation to the original Hamiltonian.  Finally, we
obtained a crude estimate of the parameters governing the transition
from coherence to incoherence in the disordered system, using a simple
Lindemann criterion which may be applicable at $T = 0$.

Viewed as a representation of a Josephson junction array
(or a thin superconducting film), our model, and our
approximations used to treat it, are both very
simplified representations of a real superconducting array.  For
example, a bimodal distribution of charging energies is
oversimplified, as is our assumption of diagonal charging energies and
diagonal charging disorder.  In addition, we have not considered
dissipation, which is known to have important effects on S/I
transitions in Josephson arrays. \cite{chakra} Nevertheless, our
results may provide some useful insights in understanding the S/I
transition in such arrays, as well as the spectrum of excitations 
to be expected in disordered arrays.

Would it be possible to measure experimentally the phase phonon
spectrum we have calculated in this paper?  It is not clear what
experiment would be directly sensitive to this spectrum in a
superconducting array or film.  A more promising direction might be
the excitation spectrum of a Bose superfluid (e.g. $^4$He) in a porous
medium.  Such spectra have been extensively studied experimentally,
primarily using inelastic neutron scattering
techniques. \cite{sokol,kinder,plantevin,anderson,dimeo}  While the
behavior of $^4$He in such porous media is certainly more complex than
the relatively simple model discussed here, some aspects of the
observed behavior (e. g., the persistence of rather sharp excitation
peaks even in highly disordered systems) seems to be mirrored in our
calculated spectral functions.  Obviously, a more refined model, aimed
specifically at the geometries of these porous glasses, is needed
before any comparison to experiment can be contemplated.

Finally, we briefly comment on the quality of the approximation
itself.  While the CPA is an excellent mean-field approach, it is
still based on a harmonic approximation to the underlying Hamiltonian.
Ideally, it would be preferable to obtain the desired spectral
functions using a more accurate approach, such as a quantum Monte
Carlo technique (see, e.g., Ref.\ \onlinecite{wallin}), or a version of the
self-consistent phonon approximation, \cite{wood,chakra} suitably
generalized to treat a disordered system.  We hope to return to such
approaches in a future publication.

\section{Acknowledgments} 

This work has been supported by the National Science Foundation, 
through grant DMR01-04987.


\begin{thebibliography}{3}
\bibitem{note} In solving the CPA equations, one must always take care
to choose the solutions which vary smoothly with frequency and which
exhibit the proper analytical properties, for example
$g_{\mathrm{CPA}}(\omega) \geq 0$.
\end{thebibliography}
\end{document}